\documentclass[pra,twocolumn,amsmath,amssymb,floatfix,reprint,footinbib,superscriptaddress,longbibliography,showkeys]{revtex4-1}
\usepackage{dcolumn}
\usepackage{graphicx}
\usepackage{mathrsfs}
\usepackage{mdwlist}
\usepackage{subfigure}
\usepackage{booktabs}
\usepackage{multirow}
\usepackage{amsmath}
\usepackage{textcomp}
\usepackage{upgreek}
\usepackage{dsfont}
\usepackage{amstext}
\usepackage{amssymb}
\usepackage{amsbsy}
\usepackage{appendix}
\usepackage{soul}
\usepackage{threeparttable}
\usepackage{bbm}
\usepackage{bm}
\usepackage{amsthm}
\usepackage{graphicx}
\usepackage{textcomp}
\usepackage{xcolor}
\usepackage{soul}
\usepackage{multirow}
\usepackage{color}
\usepackage{diagbox}
\usepackage{braket}
\usepackage[colorlinks,citecolor=blue]{hyperref}
\definecolor{Dgreen}{RGB}{0, 100, 0}
\usepackage{url}
\usepackage[colorlinks]{hyperref}
\hypersetup{%
	plainpages=true,
	breaklinks=true, 
	hypertexnames=false, 
	pageanchor=true,
	colorlinks=true,
	linkcolor={blue},
	citecolor={blue},
	urlcolor={blue},
	anchorcolor={black}
}

\begin{document}

\title{Efficient and flexible preparation of photonic NOON states in a superconducting system}

\author{Dong-Sheng Li}
\affiliation{Fujian Key Laboratory of Quantum Information and Quantum Optics, Fuzhou University, Fuzhou 350116, China}%
\affiliation{Department of Physics, Fuzhou University, Fuzhou 350116, China}%

\author{Yi-Hao Kang}\thanks{yihaokang@hznu.edu.cn}
\affiliation{School of Physics, Hangzhou Normal University, Hangzhou 311121, China}%

\author{Zhi-Cheng Shi}
\affiliation{Fujian Key Laboratory of Quantum Information and Quantum Optics, Fuzhou University, Fuzhou 350116, China}%
\affiliation{Department of Physics, Fuzhou University, Fuzhou 350116, China}%

\author{Yang Xiao}
\affiliation{Fujian Key Laboratory of Quantum Information and Quantum Optics, Fuzhou University, Fuzhou 350116, China}%
\affiliation{Department of Physics, Fuzhou University, Fuzhou 350116, China}%

\author{Ye-Hong Chen}\thanks{yehong.chen@fzu.edu.cn}
\affiliation{Fujian Key Laboratory of Quantum Information and Quantum Optics, Fuzhou University, Fuzhou 350116, China}%
\affiliation{Department of Physics, Fuzhou University, Fuzhou 350116, China}%
\affiliation{Quantum Information Physics Theory Research Team, Center for Quantum Computing, RIKEN, Wako-shi, Saitama 351-0198, Japan}%

\author{Yan Xia}\thanks{xia-208@163.com}
\affiliation{Fujian Key Laboratory of Quantum Information and Quantum Optics, Fuzhou University, Fuzhou 350116, China}%
\affiliation{Department of Physics, Fuzhou University, Fuzhou 350116, China}%

\begin{abstract}

The NOON states play a critical role as physical resources in quantum information processing and quantum metrology, yet their preparation efficiency and applicability are often constrained by complicated operational procedures or the requirement for nonlinear interactions. In this paper, we propose an efficient protocol to generate the NOON states within two microwave cavities embedded in a superconducting system, assisted by an auxiliary five-level qudit. The state preparation is accomplished in three steps for an arbitrary photon number $N$ by adjusting only external classical fields, while keeping the qudit-cavity coupling strengths and the qudit level spacings fixed. Based on parameters accessible in superconducting systems, numerical simulations show that the protocol achieves relatively high fidelity for the NOON states preparation even in the presence of parameter fluctuations and decoherence effects. Thus, this protocol may provide a practical approach for preparing the NOON states with current technology. Notably, since nonlinear interactions are not required, the protocol is flexible and has the potential to be applied across various physical systems.

\end{abstract}

\maketitle

\section{Introduction}\label{I}

Entangled states constitute a vital resource for quantum information processing (QIP) \cite{PhysRevLett.67.661,PhysRevLett.85.2392,PhysRevA.58.4394,RevModPhys.81.865,Friis2019,PhysRevA.101.012345,Li2023,202400518}. Among various types of entangled states, NOON states stand out as a particularly significant resource in both bosonic systems and atomic ensembles \cite{PhysRevLett.104.043601,PhysRevLett.105.050501,PhysRevA.96.013853}. Their importance stems not only from their dual role in verifying quantum non-locality \cite{PhysRevA.94.042119,RevModPhys.84.777,PhysRevA.65.032108} but also from their applications across diverse domains. For instance, the NOON states can enhance phase measurement sensitivity up to the Heisenberg limit, making it highly valuable in quantum precision measurements \cite{Higgins2007,1138007,PhysRevLett.98.223601,PieterKok_2004,PhysRevLett.99.070801,Giovannetti2011}. In quantum communications \cite{Gisin2007}, the NOON states are exploited as important physical resources due to the enhanced information capacity and processing efficiency. The NOON states also demonstrate strong utility in fields such as quantum metrology \cite{PhysRevA.65.052104,Mitchell2004,PhysRevLett.107.083601} and quantum optical lithography \cite{PhysRevLett.85.2733,PhysRevLett.87.013602}.

Motivated by the promising applications, numerous protocols \cite{PhysRevA.76.063808,PhysRevA.77.063826,science1188172,Merkel_2010,PhysRevLett.106.060401,PhysRevA.95.033838,Su2014,Xiong15,PhysRevA.95.022339,PhysRevA.101.033809,Kang23} have been proposed for generating the NOON states. Some earlier protocols \cite{PhysRevA.76.063808,PhysRevA.77.063826,science1188172} suggested the preparation of the NOON states using linear optical elements combined with post-selection techniques. These approaches typically operate in a probabilistic manner, as the preparation process should be restarted if the measurement outcomes are not desired. Moreover, factors such as finite transmission coefficients in the optical elements and the inefficiencies of the detectors further degrade the fidelity of the generated the NOON states \cite{PhysRevA.85.022115,PhysRevA.75.034302}.

In order to prepare the NOON states in deterministic manner, some protocols \cite{PhysRevA.95.033838,Su2014,Xiong15,PhysRevA.95.022339,PhysRevA.101.033809,Kang23} have been proposed based on unitary operations.
However, for the NOON state preparations via the unitary operations, the number of operational steps typically increases with the photon number $N$ ($N=2,3,4,\cdots$) of the target NOON states \cite{PhysRevA.95.033838,Su2014,Xiong15,PhysRevA.95.022339,PhysRevA.101.033809}. For example, in protocol \cite{PhysRevA.95.033838}, it takes $2N$ steps to prepare the NOON states with photon number $N$, where switchable and tunable couplings between the qubit and resonators are employed in each step. Similarly, in protocol \cite{Su2014}, $2N$ steps are required to generate the NOON states with photon number $N$. Although the coupling strengths are fixed during the operations in protocol \cite{Su2014}, the classical fields used to excite the qubit from a lower level to a higher level should be significantly stronger than the qubit-cavity coupling strengths in order to neglect the influence of coupling during the excitation process. By leveraging two-photon and three-photon resonances, protocol \cite{PhysRevA.101.033809} has shown an approach with reduced step number $N$ or $N-1$ for even or odd case. Nevertheless, the step number remains proportional to the photon number $N$. For the aforementioned protocols \cite{PhysRevA.95.033838,Su2014,Xiong15,PhysRevA.95.022339,PhysRevA.101.033809}, generating the NOON states with a large $N$ can lead to complicated operational procedures. Furthermore, the cumulative effects of errors and noise in extended operations may degrade the fidelity of the desired NOON states.

As a solution, unitary-operation-based NOON state generation can be achieved through fixed operational steps using nonlinear interactions \cite{PhysRevA.101.013804,Kang23,PhysRevA.111.L031301}. For instance, with the assistance of Kerr nonlinearity, a three-step protocol \cite{Kang23} has been proposed to generate the NOON states in two superconducting microwave cavities. In this protocol, although the number of steps is fixed, the reliance on Kerr nonlinearity limits the protocol's applicability, as nonlinear interactions may be challenging to implement in certain physical systems. Additionally, for different photon numbers, the control fields in protocol \cite{Kang23} are required to be individually designed, making it relatively complex to extend the protocol to cases involving a large number of photons. Therefore, it is worthwhile to develop a flexible approach for generating the NOON states using simple linear interactions.

In this paper, we propose an efficient and flexible protocol for generating the NOON states in a superconducting system consisting of two microwave cavities coupled to an auxiliary five-level qudit. The NOON states with arbitrary photon number $N$ can be generated in only three steps. In the first step, the qudit is excited from its lowest level to a superposition state of two higher levels. In the second step, effective single-photon drives, dependent on the state of the qudit, selectively displace the cavities by $\alpha_0=\sqrt{N}$ in the phase space so that the cavities evolve into coherent states with the average photon number $N$.
In the third step, frequency-matched pulses are applied to drive the qudit back to its lowest level, disentangling the qudit from the cavities and leading the two cavities evolve into the target NOON states. The impacts of the deviations of the control field and the coupling strengths, inter-cavity crosstalk, as well as the decoherence factors are studied via numerical simulation, where the results demonstrate that the protocol can produce an acceptable fidelity for the NOON states in the presence of these disturbing factors.

The advantages of the protocol can be summarized as follows. Firstly, the operational steps are fixed for preparing the NOON states with an arbitrary photon number $N$. Secondly, the control field design incorporates reverse engineering and optimal control technique, thereby enhancing robustness against potential control errors in the excitation and disentangling of the qudit. Thirdly, for different photon number $N$, only the displacement $\alpha_0$ in the second step and the modulating frequency in the third step are required to be adjusted, while the waveforms of the control fields remains unchanged. This makes the protocol easy to be extended. Fourthly, rather than complicated nonlinear interactions, only simple linear qudit-cavity couplings are utilized, so that the protocol can also be applied in various kind of physical systems. With these advantages, the protocol may provide useful perspectives in the preparation of NOON states and facilitates the NOON-state based quantum information tasks.

The paper is organized as follows. In Sec.~\ref{II}, we describe the dynamic of the entire system and derive the effective Hamiltonian for each step. In Sec.~\ref{III}, we will elaborate in detail on the dynamics and procedures for generating NOON states and the differences between the proposed approach and traditional multi-step approaches. In Sec.~\ref{IV}, we perform numerical simulations to validate the feasibility of the protocol and its robustness to systematic errors in classical control fields. We also consider the effect of the coupling strength deviations, inter-cavity crosstalk, and decoherence, including the energy relaxation and dephasing of the qudit, and the decay of the cavities. Finally, the conclusion is given in Sec.~\ref{V}.

\section{PHYSICAL MODEL} \label{II}

\begin{figure}[t]
\centering
\subfigure{\scalebox{0.4}{\includegraphics{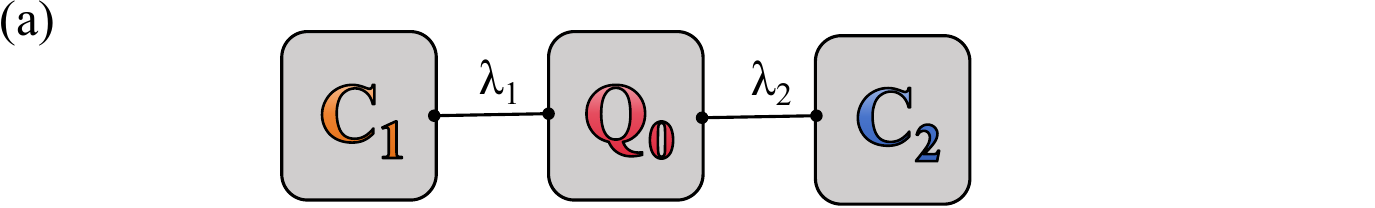}} \label{model1}}
\subfigure{\scalebox{0.34}{\includegraphics{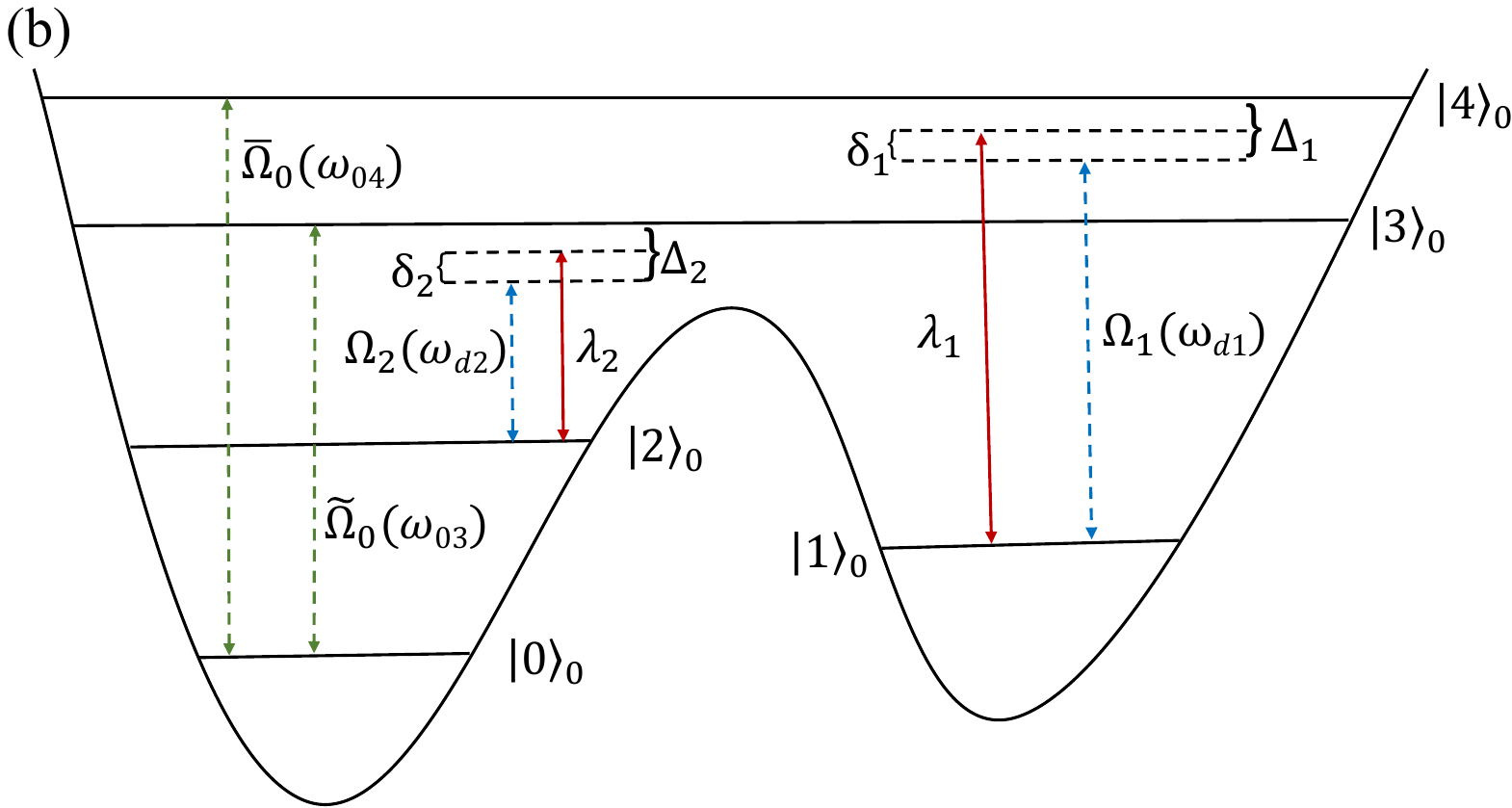}} \label{model2}}
\caption{\textcolor{blue}{(a) The physical model of the superconducting flux qudit coupling to two microwave cavities. Here, $\mathrm{C}_1$ and $\mathrm{C}_2$ are microwave cavities, and $\mathrm{Q}_0$ is a five-level qudit, and $\lambda_k$ indicates the coupling strength between the cavity and the qudit through capacitive coupling ($k=1,2$), where the cavity frequency $\omega_1=\omega_{14}-\Delta_1+\delta_1$ and $\omega_2=\omega_{23}-\Delta_2+\delta_2$. The transition $|1\rangle_0\leftrightarrow|4\rangle_0$ is driven by a classical field with the Rabi frequency (frequency) $\Omega_1$($\omega_{d1}$) and the transition $|2\rangle_0\leftrightarrow|3\rangle_0$ is driven by a classical field with the Rabi frequency (frequency) $\Omega_2$($\omega_{d2}$). (b) The energy level structure of the five-level qudit is shown below. The five energy levels of the qudit are $|0\rangle_0$, $|1\rangle_0$, $|2\rangle_0$, $|3\rangle_0$, and $|4\rangle_0$, respectively.}}
\end{figure}\label{ff1}

In this section, we introduce the physical model for generating the NOON states. As shown in Fig.~\ref{model1}, the system contains a five-level qudit (denoted by $\mathrm{Q}_0$) and two microwave cavities (denoted by $\mathrm{C}_1$ and $\mathrm{C}_2$). The energy level structure of the qudit is shown in Fig.~\ref{model2}. The  energy levels of the qudit $\mathrm{Q}_0$ is labeled as $|j\rangle_0$ $(j=0,1,2,3,4)$. The frequency of the transition $\ket{j}_0\leftrightarrow\ket{j'}_0$ $(j,j'=0,1,2,3,4)$ is denoted by $\omega_{j,j'}$, and the frequency of the cavity $\mathrm{C}_k$ ($k=1,2$) is given by $\omega_k$. The qudit is initially in the ground state $|0\rangle_0$. Each cavity is initially in the vacuum state.

The transitions $|0\rangle_0\leftrightarrow|3\rangle_0$ and $|0\rangle_0\leftrightarrow|4\rangle_0$ are resonantly driven by the classical fields with Rabi frequencies $\widetilde{\Omega}_0(t)$ and $\overline{\Omega}_0(t)$, respectively. The transition $|1\rangle_0\leftrightarrow|4\rangle_0$ is off-resonantly driven by the classical fields with Rabi frequencies $\Omega_1(t)$ and the detuning $\Delta_1$. Meanwhile, the transition $|1\rangle_0\leftrightarrow|4\rangle_0$ is coupled to the cavity $\mathrm{C}_1$ with the coupling strength $\lambda_1$ and the detuning $\omega_{14}-\omega_1=\Delta_1-\delta_1$. Analogously, the transition $|2\rangle_0\leftrightarrow|3\rangle_0$ is non-resonantly driven by the classical fields with Rabi frequencies $\Omega_2(t)$ and the detuning $\Delta_2$ and is coupled to the cavity $\mathrm{C}_2$ with the coupling strength $\lambda_2$ and the detuning $\omega_{23}-\omega_2=\Delta_2-\delta_2$. The Hamiltonian of the system under the rotating-wave approximation reads
\begin{eqnarray}\label{e1}
H&=&H_0+H_1+H_2+H_{\delta},\cr\cr
H_0&=&\widetilde{\Omega}_0(t)|0\rangle_0\langle3|+\overline{\Omega}_0(t)|0\rangle_0\langle4|+\text{H.c.}, \cr\cr
H_1&=&\Omega_1(t)e^{i\Delta_1t}|4\rangle_0\langle1|+\lambda_1a_1^{\dag}e^{-i\Delta_1t}|1\rangle_0\langle4|+\text{H.c.},\cr\cr
H_2&=&\Omega_2(t)e^{i\Delta_2t}|3\rangle_0\langle2|+\lambda_2a_2^{\dag}e^{-i\Delta_2t}|2\rangle_0\langle3|+\text{H.c.},\cr\cr
H_{\delta}&=&\delta_1a_1^{\dag}a_1+\delta_2a_2^{\dag}a_2,
\end{eqnarray}
where $a_k$ ($a_k^{\dag}$) is the annihilation (creation) operator for the cavity $\mathrm{C}_k$.

In the case that the states $\ket{1}_0$ and $\ket{2}_0$ of the qudit is initially unoccupied, under the large detuning condition $\Delta_k\gg\Omega_k,\lambda_k,\delta_k$, we obtain an effective Hamiltonian utilizing the second-order perturbation theory \cite{p07060} as
\begin{eqnarray}\label{e2}
H_{\mathrm{eff}}&=&H_0+H_1'+H_2'+H_{\delta},
\end{eqnarray}
with
\begin{eqnarray}
H_1'&=&[\frac{\lambda_1^2}{\Delta_1}a_1^{\dag}a_1+\frac{\lambda_1}{\Delta_1}(\Omega_1(t)a_1^{\dag}+\Omega_1(t)^*a_1)\cr\cr
&&+\frac{\lambda_1^2+|\Omega_1(t)|^2}{\Delta_1}]\otimes|4\rangle_0\langle4|,\cr\cr
H_2'&=&[\frac{\lambda_2^2}{\Delta_2}a_2^{\dag}a_2+\frac{\lambda_2}{\Delta_2}(\Omega_2(t)a_2^{\dag}+\Omega_2(t)^*a_2)\cr\cr
&&+\frac{\lambda_2^2+|\Omega_2(t)|^2}{\Delta_2}]\otimes|3\rangle_0\langle3|.
\end{eqnarray}
For simplicity, we consider $\Omega_1(t)\!=\!\Omega_2(t)$, $\lambda_1\!=\!\lambda_2$, $\Delta_1\!=\!\Delta_2$, $\delta_0\!=\!\lambda_1^2/\Delta_1$, and $\delta'\!=\!\delta_1\!=\!\delta_2$ in following discussion.

\section{System dynamics and Procedures for generating the NOON states}\label{III}

In this section, we will elaborate in detail on the dynamics and three steps for generating NOON states. At the beginning, both cavities are prepared in the vacuum state, and the qudit is initialized to the ground state $|0\rangle_0$. The initial state of the entire system can be represented by $\ket{\Psi(0)}=\ket{0,0,0}_{0,1,2}$, which can be written as $\ket{0,0,0}$ for short. The entire process can be divided into three steps, denoted by Step 1, Step 2, and Step 3, with the final evolution time assumed as $\tau_1$, $\tau_2$, and $\tau_3$, respectively.

In Step 1, we turn off the non-resonant driving fields $\Omega_{1}$ and $\Omega_{2}$ and turn on the resonant driving fields $\widetilde{\Omega}_0(t)$ and $\overline{\Omega}_0(t)$. Supposing $\widetilde{\Omega}_0(t)\!=\!\overline{\Omega}_0(t)\!=\!\Omega_{s_1}(t)/\sqrt{2}$, the effective Hamiltonian of the whole system is written as
\begin{eqnarray}\label{e3}
H_1^{\mathrm{eff}}(t)&=&\widetilde{\Omega}_0(t)|0,0,0\rangle\langle3,0,0|+\overline{\Omega}_0(t)|0,0,0\rangle\langle4,0,0|\cr&+&\text{H.c.}, \cr\cr
&=&\Omega_{s_1}(t)|0,0,0\rangle\langle\Phi_+|+\text{H.c.},
\end{eqnarray}
with $|\Phi_+\rangle=(|3,0,0\rangle+|4,0,0\rangle)/\sqrt{2}$. Here, the effect of the Stark-shift terms induced by off-resonant qudit-cavity couplings can be neglected under the condition $\Omega_{s_1}(t)\gg\lambda_k^2/\Delta_k$. Based on the Hamiltonian in Eq.~(\ref{e3}), we can achieve the evolution
\begin{eqnarray}\label{e4}
|\Psi(0)\rangle\!=\!|0,0,0\rangle\rightarrow|\Psi(\tau_1)\rangle=|\Phi_+\rangle,
\end{eqnarray}
using a simple resonant $\pi$-pulse, with $\Omega_{s_1}(t)\!=\!-i\pi/2\tau_1$. However, previous works~\cite{Ruschhaupt_2012,PhysRevLett.111.050404,PhysRevA.95.063403} have demonstrated the sensitivity of the resonant $\pi$-pulse to the systematic errors. As an alternative, we can use an optimized pulse designed by reverse engineering and nullification of the systematic-error-sensitivity to improve the performance of the protocol~\cite{Ruschhaupt_2012,PhysRevA.97.062317,PhysRevA.102.022617,Kang:22,PhysRevA.109.022437,202200483}. The specific expressions for the optimized pulse can be obtained as
\begin{eqnarray}
\mathrm{Re}[\Omega_{s_1}(t)]=\frac{\dot{\theta}}{2}\left(4\sin\beta\sin^3\theta-\cos\beta\right),\cr\cr
\mathrm{Im}[\Omega_{s_1}(t)]=\frac{\dot{\theta}}{2}\left(4\cos\beta\sin^3\theta+\sin\beta\right),\cr\cr
\beta(t)=\frac{4}{3}\sin^3\theta,\ \theta(t)=\pi\sin^2\left(\frac{\pi t}{2\tau_1}\right),
\end{eqnarray}
and its waveform is shown in Fig.~\ref{pulse1}. The detailed derivation process of the optimized pulse is provided in the appendix~\ref{appenA}. 

\begin{figure}
  \centering
  \includegraphics[width=8cm]{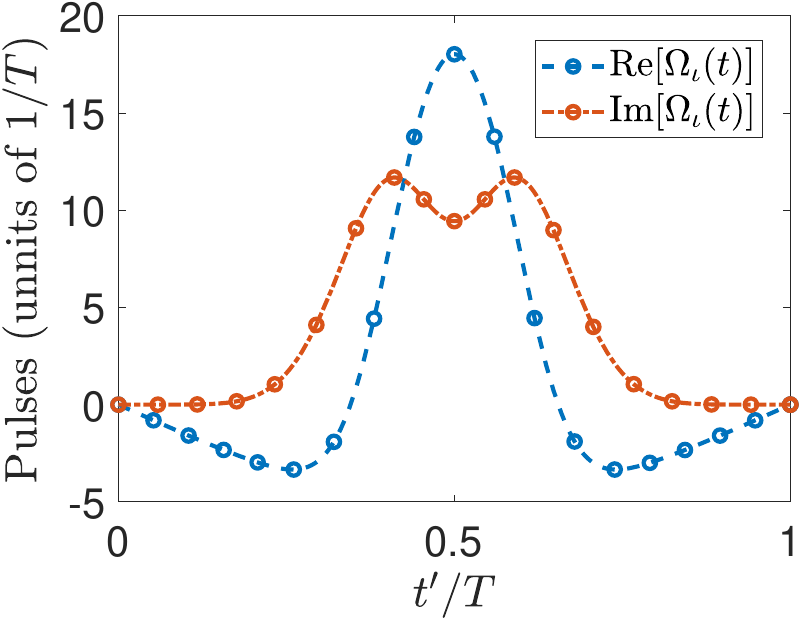}\caption{The control fields $\mathrm{Re}[\Omega_{\iota}(t)]$ and $\mathrm{Im}[\Omega_{\iota}(t)]$ versus $t'/T$. For Step 1, $\iota=s_1$, $T=\tau_1$, and $t'=t$. For Step 3, $\iota=s_3$, $T=\tau_3-\tau_2$, and $t'=t-\tau_2$.}\label{pulse1}
\end{figure}

In Step 2, we turn off the resonant classical fields $\widetilde{\Omega}_0(t)$ and $\overline{\Omega}_0(t)$ and turn on the off-resonant classical fields $\Omega_1(t)$ and $\Omega_2(t)$. To lead the cavities $\mathrm{C}_k$ to selectively evolve to coherent states with an average photon number $N$ when the qudit is in the state $\ket{5-k}_0$, we select
\begin{eqnarray}
\Omega_1(t)=\Omega_2(t)=i\Omega_{s_2}e^{-i\omega_{s_2}t},
\end{eqnarray}
where $\omega_{s_2}=\delta_0+\delta'$, $\Omega_{s_2}=\alpha_0\Delta_1/\lambda_1\tau_2$, $\alpha_0=\sqrt{N}$, and $\tau_2=2\pi/\omega_{s_2}$.
Since the initial state in Step 2 is $\ket{\Psi(\tau_1)}=|\Phi_+\rangle$, according to the expression of Eq.~(\ref{e2}), the effective Hamiltonian of the Step 2 can be rewritten by
\begin{eqnarray}\label{e23}
H_2^{\mathrm{eff}}&=&\sum_{k=1}^2[\omega_{s_2}a_k^{\dag}a_k+\frac{i\alpha_0}{\tau_2}(e^{-i\omega_{s_2}t}a_k^{\dag}-e^{i\omega_{s_2}t}a_k)\cr\cr
&&+\frac{\lambda_k^2+|\Omega_{s_2}|^2}{\Delta_k}]\otimes|5-k\rangle_0\langle 5-k|.
\end{eqnarray}
Governed by the Hamiltonian in Eq.~(\ref{e23}), the evolution of the system is
\begin{eqnarray}
|\Psi(\tau_1)\rangle&=&\frac{1}{\sqrt{2}}(|3,0,0\rangle+|4,0,0\rangle)\cr\cr
\rightarrow|\Psi(\tau_2)\rangle&=&\frac{e^{i\Theta_{s_2}}}{\sqrt{2}}(|3,0,\widetilde{0}\rangle+|4,\widetilde{0},0\rangle),
\end{eqnarray}
where $|\widetilde{0}\rangle=D(\alpha_0)|0\rangle$, $D(\alpha_0)=\exp[\alpha_0(a^{\dag}-a)]$, and $\Theta_{s_2}=(\lambda_1^2+|\Omega_{s_2}|)\tau_2/\Delta_1$. 

In Step 3, we turn on the resonant classical fields $\widetilde{\Omega}_0(t)$ and $\overline{\Omega}_0(t)$ as well as the off-resonant classical fields $\Omega_1(t)$ and $\Omega_2(t)$. In this case, the evolution of the system is governed
by the Hamiltonian in Eq.~(\ref{e2}). In order for the two cavities to evolve into the NOON states and the qudit to return to the initial state, it is necessary to design the specific frequency-matched classical fields. Specifically, the Rabi frequencies of the resonant driving field and the non-resonant driving field are respectively set as
\begin{eqnarray}
\widetilde{\Omega}_0(t)&=&\overline{\Omega}_0(t)=\widetilde{\Omega}_{s_3}(t)=\Omega_{s_3}(t)e^{-i(N\delta'+\widetilde{\delta})t}/\varepsilon_{N,0},\cr\cr
\Omega_1(t)&=&\Omega_{2}(t)=\Omega'_{s_3}=-\frac{\omega_{s_2}\alpha_0\Delta_1}{\lambda_1},
\end{eqnarray}
with $\widetilde{\delta}=N\omega_{s_2}\!-\!|\Omega_{s_3}|^2/\Delta_1\!-\!\delta_0$, and $\varepsilon_{N,0}=(N/e)^{N/2}/\sqrt{N!}$.
Under the above conditions, the effective Hamiltonian in Eq.~(\ref{e2}) of the system can be rewritten as
\begin{eqnarray}\label{e11}
H_3^{\mathrm{eff}}&=&H^{(1)}_{s_3}+H^{(2)}_{s_3},
\end{eqnarray}
where
\begin{eqnarray}
H^{(1)}_{s_3}
&=&\sum_{k=1}^2[\omega_{s_2}a_k^{\dag}a_k-\omega_{s_2}(a_k^{\dag}+a_k)
+\delta_0+\frac{|\Omega'_{s_3}|^2}{\Delta_k}]\cr\cr
&&\otimes|5-k\rangle_0\langle 5-k|,\cr\cr
&=&\sum_{k=1}^2[\omega_{s_2}\widetilde{a}_k^{\dag}\widetilde{a}_k
+\delta_0+\frac{|\Omega'_{s_3}|^2}{\Delta_k}-N\omega_{s_2}]\cr\cr&&\otimes|5-k\rangle_0\langle 5-k|,\cr\cr
&=&\sum_{k=1}^2[\omega_{s_2}\left(\sum_{n=0}^\infty n|\widetilde{n}\rangle_k\langle \widetilde{n}|-\widetilde{\delta}\right)]\cr\cr&&\otimes|5-k\rangle_0\langle 5-k|,
\end{eqnarray}
with $\alpha_0=\sqrt{N}$, $\widetilde{a}_k=a_k-\alpha_0=D(\alpha_0)a_kD^{\dag}(\alpha_0)$, $|\widetilde{n}\rangle_k=D(\alpha_0)|n\rangle_k$, and $\widetilde{a}_k^{\dag}\widetilde{a}_k|\widetilde{n}\rangle_k=n|\widetilde{n}\rangle_k$. The expression of the effective Hamiltonian $H^{(2)}_{s_3}$ in Eq.~(\ref{e11}) is
\begin{widetext}
\begin{eqnarray}
H^{(2)}_{s_3}&=&\sum_{k=1}^{2}\delta'a_k^{\dag}a_k|0\rangle_0\langle0|+\left(\widetilde{\Omega}_{s_3}(t)|0\rangle_0\langle5-k|+\text{H.c.}\right),\cr\cr
&=&\sum_{k=1}^{2}\delta'a_k^{\dag}a_k|0\rangle_0\langle0|+\sum_{n=0}^{\infty}\left(\widetilde{\Omega}_{s_3}(t)|0\rangle_0\langle5-k|\otimes|\widetilde{n}\rangle_k\langle \widetilde{n}|+\text{H.c.}\right),\cr\cr
&=&\sum_{k=1}^{2}\delta'a_k^{\dag}a_k|0\rangle_0\langle0|+\sum_{n,m=0}^{\infty}\left(\widetilde{\Omega}_{s_3}(t)\varepsilon_{m,n}|0\rangle_0\langle5-k|\otimes|m\rangle_k\langle \widetilde{0}|+\text{H.c.}\right),
\end{eqnarray}
where $|\widetilde{n}\rangle_k\!=\!\sum_{m=0}^{\infty}\varepsilon_{m,n}|m\rangle_k$ and $\varepsilon_{m,n}$ is the expansion coefficient.
Assuming $\omega_{s_2},\delta'\gg\Omega_{s_3}(t)$ and transforming the Hamiltonian $H_3^{\mathrm{eff}}$ in Eq.~(\ref{e11}) into a frame $\mathfrak{R}$ by using a unitary operator
\begin{equation}
R=\exp\left[-i\left(\sum_{k=1}^{2}\delta'a_k^{\dag}a_k|0\rangle_0\langle0|+H^{(1)}_{s_3}\right)t\right],
\end{equation}
the Hamiltonian $H_{3}^{\mathrm{eff}}$ in Eq.~(\ref{e11}) becomes
\begin{eqnarray}\label{e15}
\mathcal{H}_{3}^{\mathrm{eff}}=R^{\dag}H_{3}^{\mathrm{eff}}R
=\sum_{k=1}^{2}\sum_{n,m=0}^{\infty}\left(\frac{\Omega_{s_3}(t)\varepsilon_{m,n}}{\varepsilon_{N,0}}e^{i[(m-N)\delta'-n\omega_{s_2}]t}|0\rangle_0\langle5\!-\!k|\!\otimes\!|m\rangle_k\langle \widetilde{n}|+\text{H.c.}\right).
\end{eqnarray}
\end{widetext}
We only retain the terms corresponding to $m=N$ and $n=0$ while omit other high-frequency terms under the condition $\omega_{s_2},\delta'\gg\Omega_{s_3}(t)$. Then, the Hamiltonian $\mathcal{H}_{3}^{\mathrm{eff}}$ is further reduced to
\begin{eqnarray}\label{e28}
\mathcal{H}_{3}^{\mathrm{eff}}=\sum_{k=1}^{2}\left(\Omega_{s_3}(t)|0\rangle_0\langle5\!-\!k|\!\otimes\!|N\rangle_k\langle \widetilde{0}|+\text{H.c.}\right).
\end{eqnarray}

Similar to Step 1, one can realize the transformations
\begin{eqnarray}
|3,0,\widetilde{0}\rangle\rightarrow|0,0,N\rangle,\cr\cr
|4,\widetilde{0},0\rangle\rightarrow|0,N,0\rangle,
\end{eqnarray}
using a resonant $\pi$-pulse $\Omega_{s_3}(t)=i\pi/[2(\tau_3-\tau_2)]$ or using an improved pulse design as
\begin{eqnarray}
&&\mathrm{Re}[\Omega_{s_3}(t)]=\frac{\dot{\theta}}{2}\left(4\sin\beta\sin^3\theta-\cos\beta\right),\cr\cr
&&\mathrm{Im}[\Omega_{s_3}(t)]=\frac{\dot{\theta}}{2}\left(4\cos\beta\sin^3\theta+\sin\beta\right),\cr\cr
&&\beta(t)=\frac{4}{3}\sin^3\theta,\ \theta(t)=\pi\sin^2\left[\frac{\pi(t-\tau_2)}{2(\tau_3-\tau_2)}\right],
\end{eqnarray}
whose waveforms are shown in Fig.~\ref{pulse1}. In this way, the evolution of the system in the Step 3 is
\begin{eqnarray}
|\Psi(\tau_2)\rangle&=&\frac{e^{i\Theta_{s_2}}}{\sqrt{2}}(|3,0,\widetilde{0}\rangle+|4,\widetilde{0},0\rangle)\cr\cr
\rightarrow|\Psi(\tau_3)\rangle&=&\frac{e^{i(\Theta_{s_2}+4\delta'\tau_3)}}{\sqrt{2}}(|0,N,0\rangle+|0,0,N\rangle),
\end{eqnarray}
where the final state $|\Psi(\tau_3)\rangle$ of the system is the target NOON states of the two cavities, and $(\Theta_{s_2}+4\delta'\tau_3)$ is a global phase acquired during the evolutionary process.

\begin{figure}
  \centering
  \includegraphics[width=8cm]{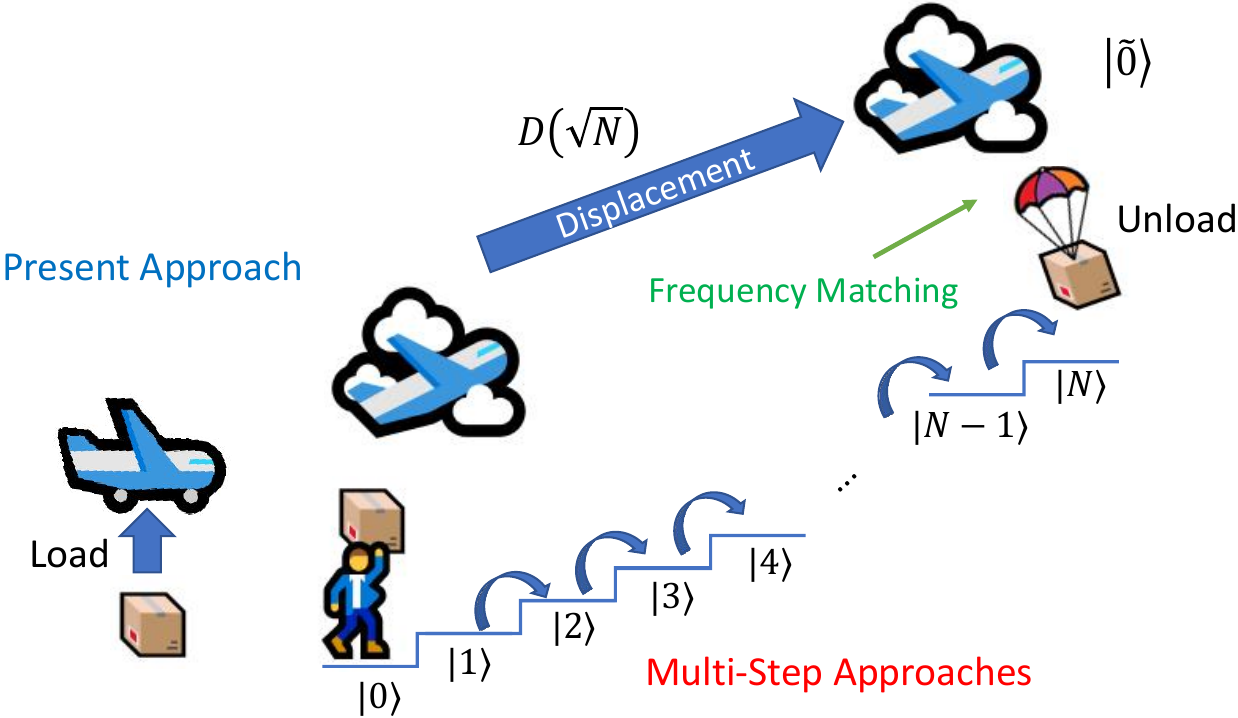}\caption{Schematic diagram of NOON states preparation, where $|\widetilde{0}\rangle=D(\sqrt{N})|0\rangle$ is the displacement Fock state, and $|n\rangle$ ($n=1,2,\dots,N$) is the Fock state.}\label{approach}
\end{figure}

An analogy of the proposed approach is demonstrated in Fig.~\ref{approach}. The higher levels $\ket{3}_0$ and $\ket{4}_0$ play the role of the aircraft in the figure. While the lowest level $\ket{0}_0$ acts as the floors. The photon number $n$ ($n=0,1,2,...$) is analogized with the hight of the floor, and the state of the cavities is deemed as the goods to be transported. In the first step, as the classical fields $\Omega_1(t)$ and $\Omega_2(t)$ are turned off, the state of the cavities remain in the vacuum state under the dispersive couplings. The excitation of the qudit to the higher levels $\ket{3}_0$ and $\ket{4}_0$ is similar to load the goods on the aircraft. In the second step, we turn on the classical fields $\Omega_1(t)$ and $\Omega_2(t)$, and the cavity modes are selectively displaced to the coherence state $\ket{\widetilde{0}}$ with average photon number equal to $N$. This is similar to the scenario that the aircraft takes off and bring the goods to the airspace above the destination. In the last step, we use the frequency-matched fields to drive the qudit to its lowest level, and the system finally populate the target NOON states with photon number $N$. The frequency-matched field acts as an parachute, which unload the goods safely to the destination.

For the conventional multi-step approaches, the photon number of cavities are raised one by one by repeating certain procedures, which is similar to climbing a Fock-state ladder as shown by the porter in Fig.~\ref{approach}. When the target photon number $N$ increases, the operational steps should also be increase. However, for the proposed approach, we only need to adjust the displacement $\alpha_0=\sqrt{N}$ in the second step, while keep the total operational steps unchanged. This improve the efficiency of the NOON states preparation with a relatively large photon number $N$.

\section{NUMERICAL SIMULATIONS AND DISCUSSIONS}\label{IV}

In this section, numerical simulations are performed to check the feasibility and robustness of the protocol. The initial state of the system is prepared in
\begin{eqnarray}
|\Psi(0)\rangle\!=\!|0,0,0\rangle.
\end{eqnarray}
The feasibility and robustness of the protocol are discussed in detail in three parts below.

\subsection{Verification of the protocol with different photon numbers}
\begin{figure*}
  \centering
  \includegraphics[width=\linewidth]{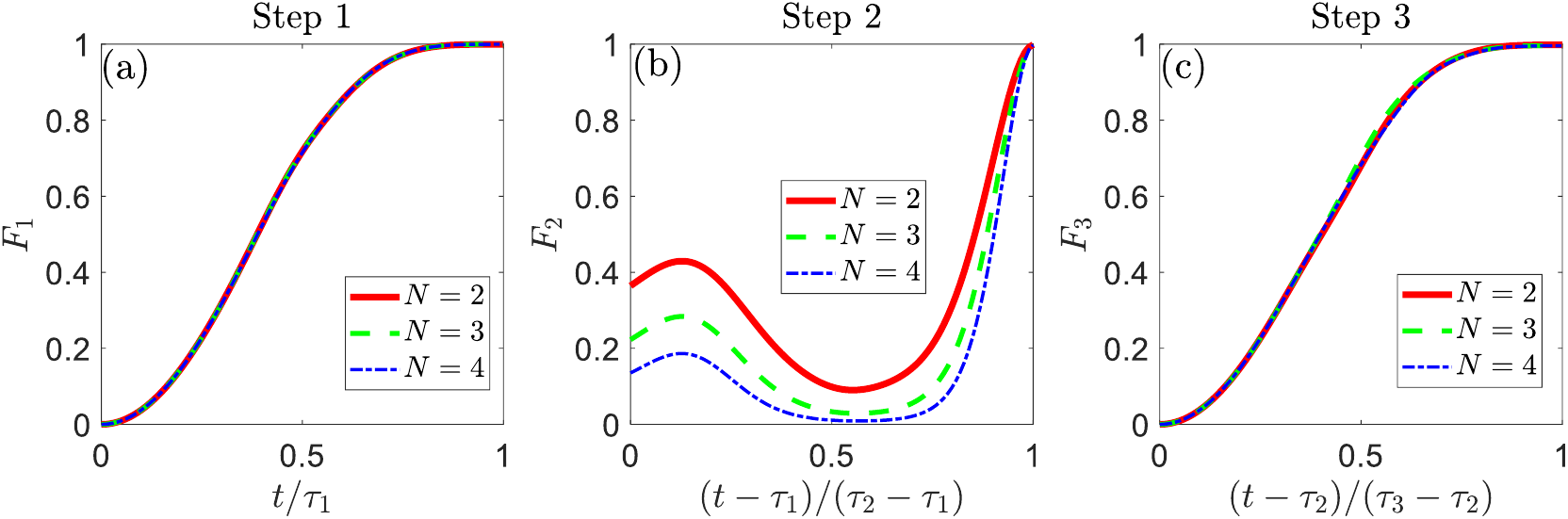}\caption{\textcolor{blue}{Fidelity $F_p$ of the target state in Step $p$ ($p=1,2,3$) versus $t$ for the photon number $N=2,3,4$.}}\label{output1}
\end{figure*}

\textcolor{blue}{To validate the above theoretical analysis and feasibility of the protocol, we investigated the evolution of system under the original Hamiltonian $H$ in Eq.~(\ref{e1}) and the effective Hamiltonian $H_e$, where $H_e=H_1^{\mathrm{eff}}, H_2^{\mathrm{eff}}, \mathcal{H}_{3}^{\mathrm{eff}}$ for Step 1,2,3, respectively.~When the system is controlled by the original Hamiltonian $H$, the evolution is governed by the equation
\begin{eqnarray}\label{e32}
\dot{\rho}(t)&=&-i[H,\rho(t)],
\end{eqnarray}
where $\rho(t)$ is the density operator of the total system. To verify the theoretical analysis and feasibility of the protocol, we consider the cases of photon numbers $N=2,3,4$ as examples.} The fidelity of the target state in each step is defined as~\cite{PhysRevA.101.032329,PhysRevA.105.042434,PhysRevApplied.23.054037}
\begin{eqnarray}
F_p=\sqrt{\langle\Psi(\tau_p)|\rho(t)|\Psi(\tau_p)\rangle},
\end{eqnarray}
where $p=1,2,3$, and
\begin{eqnarray}
|\Psi(\tau_1)\rangle&=&|\Phi_+\rangle=\frac{1}{\sqrt{2}}(|3,0,0\rangle+|4,0,0\rangle), \cr\cr
|\Psi(\tau_2)\rangle&=&\frac{e^{i\Theta_{s_2}}}{\sqrt{2}}(|3,0,\widetilde{0}\rangle+|4,\widetilde{0},0\rangle),\cr\cr
|\Psi(\tau_3)\rangle&=&\frac{e^{i(\Theta_{s_2}+4\delta'\tau_3)}}{\sqrt{2}}(|0,N,0\rangle+|0,0,N\rangle).
\end{eqnarray}
It is worth noting that $|\Psi(\tau_3)\rangle$ is the desired NOON states and $F_3$ is its corresponding fidelity.
The five-level flux qudit is selected to implement the protocol, and the corresponding frequencies of five energy levels $|0\rangle_0$, $|1\rangle_0$, $|2\rangle_0$, $|3\rangle_0$, and $|4\rangle_0$ are $0$, $2\pi\times3$ GHz, $2\pi\times5$ GHz, $2\pi\times15$ GHz, and $2\pi\times20$ GHz, respectively \cite{PhysRevA.105.042434,PhysRevB.86.140508,PhysRevLett.115.223603}. The frequencies $\omega_1$ and $\omega_2$ of the cavities $\mathrm{C}_1$ and $\mathrm{C}_2$ are $2\pi\times11.0346$ GHz and $2\pi\times4.0346$ GHz, respectively \cite{PhysRevA.105.042434}. \textcolor{blue}{In order to compare the differences between the evolution governed by the effective Hamiltonian $H_e$ and the original Hamiltonian $H$, the evolution leaded by the effective Hamiltonian $H_e$ is discussed in appendix \ref{appenB}. When the system is controlled by the effective Hamiltonian $H_e$, we can precisely obtain the target NOON states (the fidelity $F_3$ can reach 1 at final moment). This confirms the effectiveness of the above theoretical analysis. For the photon number $N=2,3,4$, the fidelity $F_p$ of the target state in each step versus $t$ is plotted in Fig.~\ref{output1} and the corresponding parameters are presented in Table I, where the total evolution time of the system is $T_f=15~\mu\mathrm{s}$. As illustrated in Figs.~\ref{output1}(a), (b), and (c), the fidelity $F_1$ ($F_2$, $F_3$) surpasses 0.999 (0.99, 0.99) after Step 1 (2, 3) for different photon numbers. This indicates that the evolution of the system in each step follows the evolution predicted by the effective Hamiltonian. It should be noted that the initial states of the second and third steps in the evolution here are both derived from the final states of the first and second steps, respectively, which ensures the continuity of the system evolution. From the Step 2 of the system evolution (Figs.~\ref{output1}(b)), it can be seen that the fidelity $F_2$ is not zero at the initial moment for different photon numbers. This is because the initial state $|\Psi(\tau_1)\rangle$ in Step 2 and the corresponding target state $|\Psi(\tau_2)\rangle$ are not completely orthogonal.} \textcolor{blue}{In addition, for clarity, the fidelity $F_3$ of the target NOON states versus $t$ throughout the evolution process is plotted in Fig.~5. As shown Fig.~5, vertical dashed lines are used to mark the evolution duration of each step. Additionally, we clearly display the fidelity $F_3$ of the NOON states at the final moment through the subfigure at the top of Fig.~5. Obviously, the fidelity $F_3$ is over 0.99 when the three steps are completed for the photon number $N=2,3,4$. This also indicates that, the protocol can be successfully applied to the generation of the NOON states with different photon numbers. In the following discussions, for simplicity, we consider the photon number $N=4$ and the variations of the fidelity $F_3$ throughout the evolution process when estimating the performance of the protocol under the influence of various disturbing factors.}

\begin{figure}
  \centering
  \includegraphics[width=8cm]{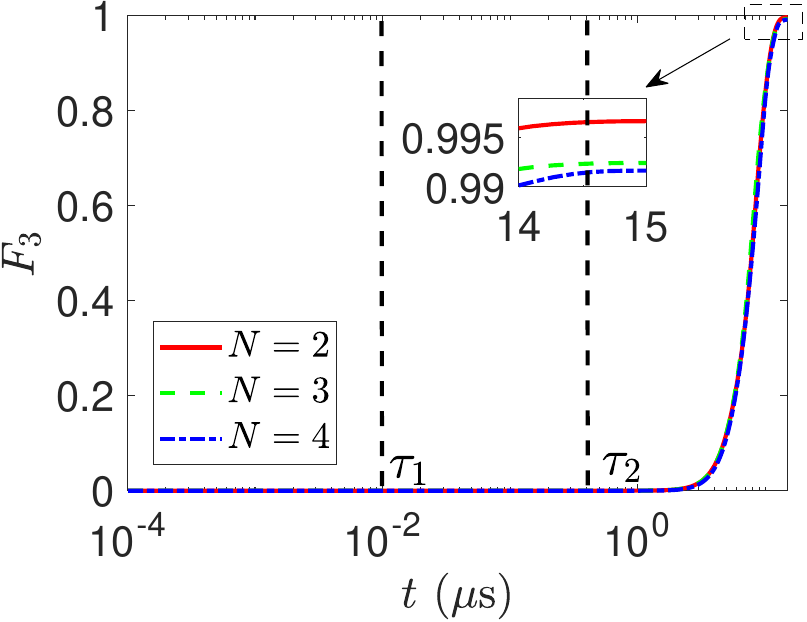}\caption{\textcolor{blue}{Fidelity $F_3$ versus $t$ for the photon number $N=2,3,4$, where $\tau_1=0.01 \mu$s, and $\tau_2=0.31 \mu$s.}}\label{NOON234}
\end{figure}

\begin{table}\centering
\caption{The values of parameters $\delta'$, $\Delta_k$, and $\lambda_k$ for the photon numbers $N=2,3,4$, $k=1,2$.}\label{tab}
{\begin{tabular}{cccc} \hline\hline
&&\\[-7pt]
\ \ $N$\ \ \ & \ \ $\delta'/2\pi$ (MHz)\ \ & \ \ $\Delta_k/2\pi$ (GHz)\ \  & \ \ $\lambda_k/2\pi$ (MHz)\ \ \\[1pt]
\hline
&&\\[-7pt]
\ \ $2$ \ \ \ &\ \ $-0.67$\ \ & \ \ $5.00$\ \ & \ \ $141.42$\ \    \\
&&\\[-4pt]
\ \ $3$ \ \ \ &\ \ $-0.67$\ \ & \ \ $7.50$\ \ & \ \  $173.48$\ \
\\
&&\\[-4pt]
\ \ $4$ \ \ \ &\ \ $-0.67$\ \ & \ \ $5.96$\ \ & \ \  $130.00$\
\\[2pt]
\hline \hline
\end{tabular}}
\end{table}

\subsection{Effects of systematic errors on the protocol}\label{Ab}

\begin{figure}
  \centering
  \includegraphics[width=8cm]{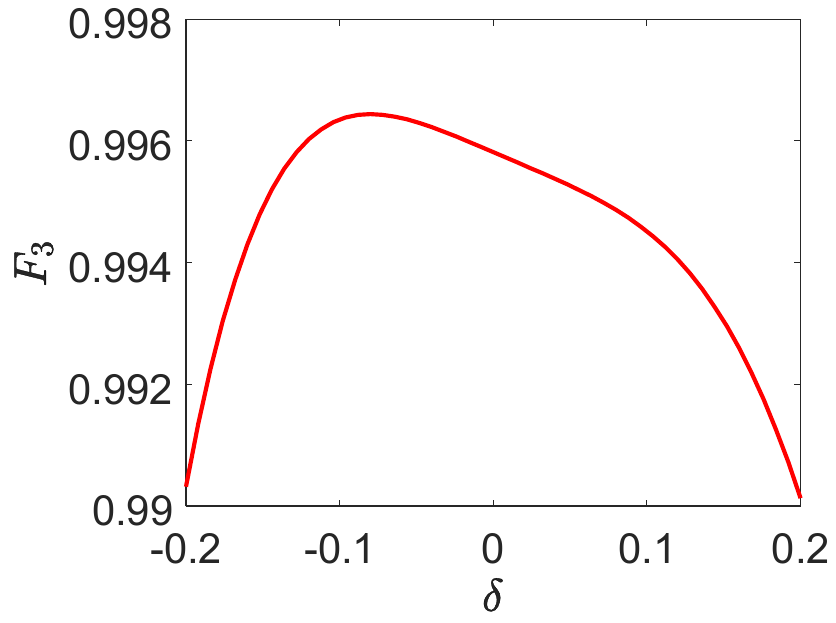}\caption{Fidelity $F_3$ versus the systematic error rate $\delta$ for $N=4$.}\label{epsilont}
\end{figure}

Due to imperfections in experimental instrumentation and operations, systematic errors are inherent in practical quantum implementations \cite{Ruschhaupt_2012,PhysRevLett.111.050404,KYHPRA111_2025}. In the present protocol, there may exist systematic errors in the strengths of the Rabi frequencies and the coupling strengths. We will separately discuss the impacts of these errors. First, let us consider the systematic errors of the Rabi frequencies $\widetilde{\Omega}_0(t)$ and $\overline{\Omega}_0(t)$, which become $(1+\delta)\widetilde{\Omega}_0(t)$ and $(1+\delta)\overline{\Omega}_0(t)$ in the presence of the systematic errors, respectively. Here, $\delta$ denotes the error rate. The fidelity $F_3$ versus the error rate $\delta$ is plotted in Fig.~\ref{epsilont}. On the whole, the fidelity $F_3$ decreases as $|\delta|$ increases for the error rate $\delta\in[-0.2,0.2]$~\cite{PhysRevA.107.013702,PhysRevA.109.042615,Zhao2021}. However, the value of $F_3$ is always greater than $99\%$ when $\delta\in[-0.2,0.2]$. This demonstrates that the protocol with the optimized pulse is robust to the systematic errors of the Rabi frequencies $\widetilde{\Omega}_0(t)$ and $\overline{\Omega}_0(t)$. It should be noted that the highest point of fidelity is not the point where the systematic error $\delta$ is zero. This is because we use the second-order perturbation theory to calculate the effective Hamiltonian in Sec.~\ref{II}, which may also lead to errors as the terms with higher orders are omitted.

\begin{figure}[tbhp]
  \centering
  \includegraphics[width=8cm]{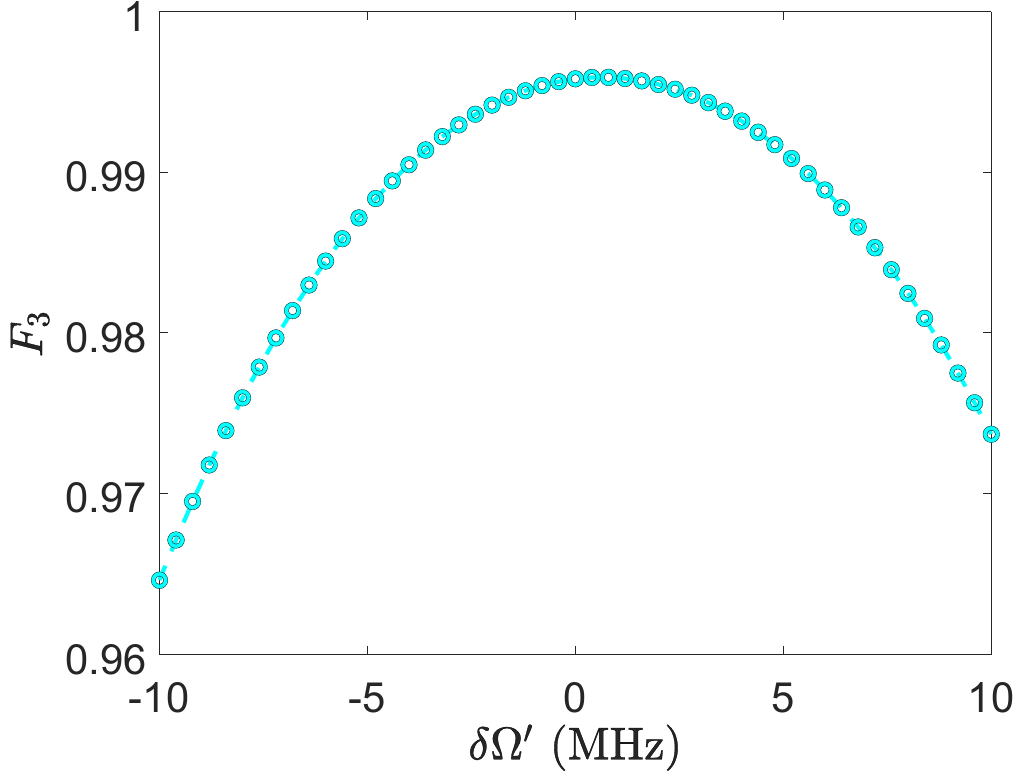}\caption{Fidelity $F_3$ versus the deviation $\delta\Omega'$ of Rabi frequencies $\Omega_1(t)$ and $\Omega_2(t)$.}\label{deltaomega}
\end{figure}

Next, we investigate systematic errors of the Rabi frequencies $\Omega_1(t)$ and $\Omega_2(t)$. In this case, the Rabi frequencies $\Omega_1(t)$ and $\Omega_2(t)$ become $\Omega_1(t)+\delta\Omega'$ and $\Omega_2(t)+\delta\Omega'$, where $\delta\Omega'$ is the deviation of Rabi frequency. In Fig.~\ref{deltaomega}, we plot the fidelity $F_3$ versus the deviation $\delta\Omega'$ in the range $[-10,10]~\mathrm{MHz}$. As shown in Fig.~\ref{deltaomega}, the fidelity $F_3$ decreases as $|\delta\Omega'|$ increases. The value of $F_3$ still exceeds $96\%$ when $\delta\Omega'$ reaches $-10\ \mathrm{MHz}$. This shows that the protocol can still generate NOON states with acceptable fidelity when considering the systematic errors in the Rabi frequencies $\Omega_1(t)$ and $\Omega_2(t)$.

\begin{figure}[thbp]
  \centering
  \includegraphics[width=8cm]{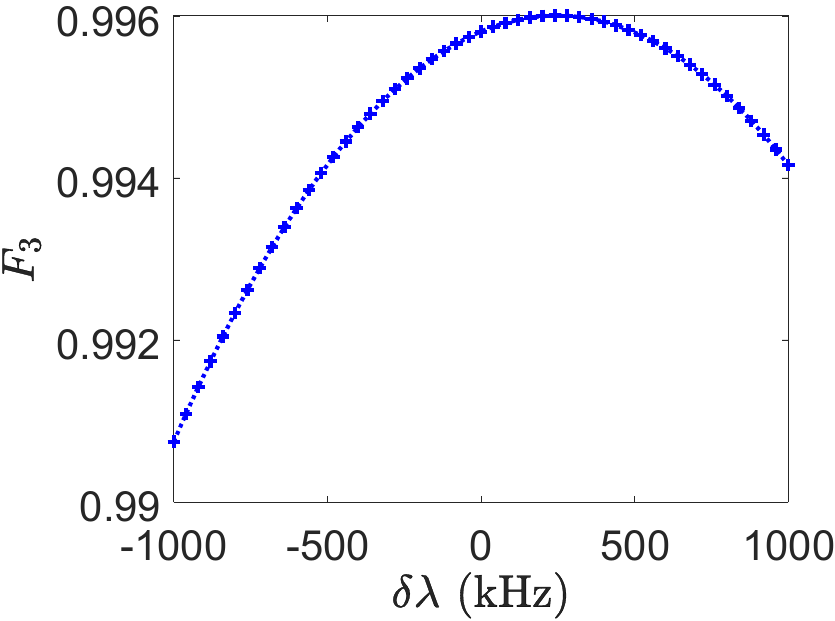}\caption{Fidelity $F_3$ versus the drift $\delta\lambda$ of the coupling strengths $\lambda_1$ and $\lambda_2$.}\label{deltalambda}
\end{figure}

At last, we estimate the influence of the deviation of the coupling strengths $\lambda_1$ and $\lambda_2$, where the coupling strengths change as $\lambda_1\rightarrow\lambda_1+\delta\lambda$, $\lambda_2\rightarrow\lambda_2+\delta\lambda$. Here, $\delta\lambda$ is the strength drift of the coupling. The fidelity $F_3$ versus the strength drift is demonstrated in Fig.~\ref{deltalambda}. As shown by the figure, the fidelity $F_3$ decreases as $|\delta\lambda|$ increases. This is because the systematic errors of the coupling strengths will cause a frequency mismatch in the effective Hamiltonian in Eq.~(\ref{e23}) and Eq.~(\ref{e15}), which destroys the desired evolution of the system. This suggests that the coupling strengths should be more precisely control during the generation of the NOON states.

\subsection{Effects of inter-cavity crosstalk on the protocol}

\begin{figure}
  \centering
  \includegraphics[width=8cm]{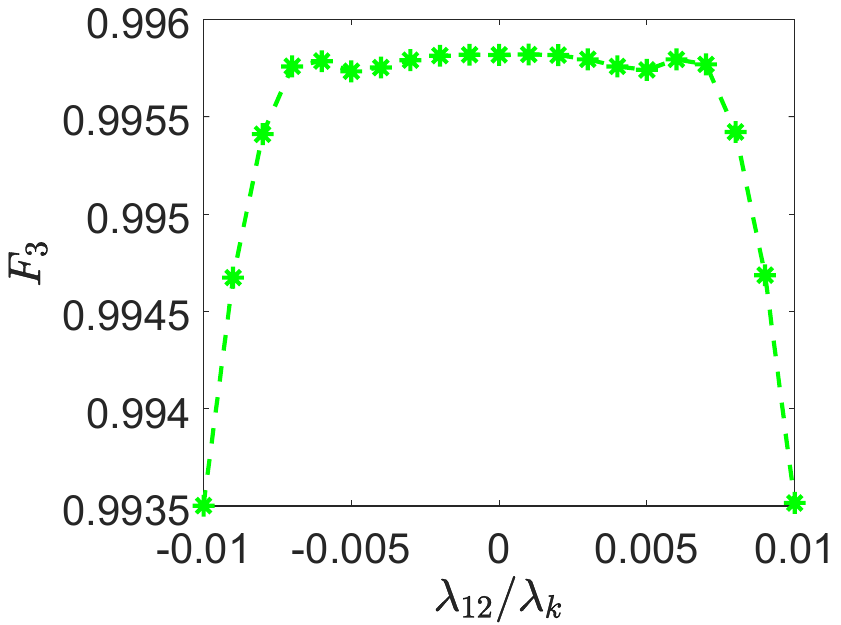}\caption{Fidelity $F_3$ versus versus the crosstalk coupling strengths $\lambda_{12}/\lambda_k$.
  }\label{crosstalk}
\end{figure}

In the experimental implementation, the crosstalk between the cavities are inevitable \cite{PhysRevA.93.042307}, which will affect the evolution of the system. Here, the inter-cavity crosstalk is described as
\begin{eqnarray}\label{36}
H_{cr}&=&\lambda_{12}a_1^{\dag}a_2e^{i\Delta't}+\text{H.c.},
\end{eqnarray}
where $\lambda_{12}$ is the crosstalk coupling strengths, and $\Delta'=|\omega_1-\omega_2|$ is the detuning between both cavities. When the crosstalk is taken into account, the Hamiltonian of the system is $H_r=H+H_{cr}$, and the evolution of the system can be described by the equation
\begin{eqnarray}
\dot{\rho}(t)&=&-i[H_{r}, \rho(t)].
\end{eqnarray}
The crosstalk coupling strengths $\lambda_{12}\in[-0.01\lambda_k,0.01\lambda_k]$ \cite{PhysRevA.76.042319,PhysRevA.105.062436} is selected to test the performance of the protocol. We plot the fidelity $F_3$ versus the coupling strengths $\lambda_{12}/\lambda_k$ in Fig.~\ref{crosstalk}. As shown in Fig.~\ref{crosstalk}, overall, the fidelity $F_3$ decreases with the increase of the crosstalk coupling strengths. However, for $\lambda_{12}/\lambda_k\in[-0.01,0.01]$ ($k=1,2$), the values of $F_3$ are always greater than $99.35\%$. This shows that the fidelity of the protocol can maintain a relatively high level under the influence of current crosstalk. In fact, this result is predictable because the inter-cavity crosstalk $H_{cr}$ can be omitted as the high-frequency term due to the detuning $\Delta'$ is much greater than the crosstalk coupling strength $\lambda_{12}$.

\subsection{Effects of decoherence on the protocol}\label{Ac}

\setlength{\tabcolsep}{6pt}
\begin{table*}[t]
\caption{Samples of the fidelity $F_3$ with multiple disturbing factors.}\label{tab2}
{\begin{tabular}{cccccccc} \hline\hline
&&\\[-7pt]
$\delta$  & \ $\delta\Omega' $ (MHz) & $\delta\lambda $ (kHz) & \ $\lambda_{12}$  &  $\Gamma_d $ (kHz) &  \ $\Gamma_r $ (kHz) &  $\Gamma_{\kappa} $ (kHz) & $F_3(T)$     \\[1pt]
\hline &&\\[-7pt]
$0.01$  &  $1$ &  $100$ &  $0.001$  &  $5$  & $5$  & $1$  & $94.99\%$  \\
&&\\[-4pt]
$0.02$  &  $5$ &  $500$ &  $0.005$  &  $10$  & $10$  & $1$  &  $93.61\%$  \\
&&\\[-4pt]
$0.02$ & $5$ &  $500$ &  $0.01$  & $15$ & $15$  & $1$    &   $90.38\%$\\
&&\\[-10pt]
\hline \hline
\end{tabular}}
\end{table*}

Since the system cannot be completely isolated from the environment, the influence of environment-induced decoherence should also be taken into account. The main decoherence factors in the protocol are the qudit dephasing on the levels $|3\rangle_0$ and $|4\rangle_0$, the qudit energy relaxation of the paths $|4\rangle_0\rightarrow|i\rangle_0$, $|3\rangle_0\rightarrow|j\rangle_0$, $|2\rangle_0\rightarrow|0\rangle_0$, $|1\rangle_0\rightarrow|0\rangle_0$ ($i=0,1,2,3,j=0,1,2$), and the single photon loss of each cavity. The evolution of the system in the presence of decoherence can be described by the Lindblad master equation \cite{10.1063/5.0134394,KYHSCPMA68_2025}
\begin{eqnarray}\label{mas}
\dot{\rho}(t)&=&-i\left[H(t),\rho(t)\right] + \sum_{f=3,4}\mathcal{\gamma}_{ff}\mathcal{L}[\sigma_{kk}]\rho(t)\cr\cr
&+&\sum_{i=0}^{3}\mathcal{\gamma}_{4i}\mathcal{L}[\sigma_{4i}^-]\rho(t)
+\sum_{j=0}^{2}\mathcal{\gamma}_{3j}\mathcal{L}[\sigma_{3j}^-]\rho(t)\cr\cr
&+&\mathcal{\gamma}_{20}\mathcal{L}[\sigma_{20}^-]\rho(t)+\mathcal{\gamma}_{10}\mathcal{L}[\sigma_{10}^-]\rho(t)\cr\cr
&+&\sum_{k=1}^2\kappa_k\mathcal{L}[a_k]\rho(t),
\end{eqnarray}
with $\sigma_{ff}=|f\rangle_0\langle f|$ ($f\!=\!3,4$), $\sigma_{4i}^-=|4\rangle_0\langle i|$, $\sigma_{3j}^-=|3\rangle_0\langle j|$, $\sigma_{20}^-=|2\rangle_0\langle 0|$, and $\sigma_{10}^-=|1\rangle_0\langle 0|$. In addition, $\mathcal{L}[O]$ is the Lindblad superoperator satisfies $\mathcal{L}[O]\rho(t)=O\rho(t)O^{\dag}-O^{\dag}O\rho(t)/2-\rho(t)O^{\dag}O/2$ with $O=\sigma_{ff},\sigma_{4i}^-,\sigma_{3j}^-,\sigma_{20}^-,\sigma_{10}^-,a_n$. Here, $\mathcal{\gamma}_{ff}$ is the dephasing rate, $\mathcal{\gamma}_{4i}$ ($\mathcal{\gamma}_{3j}$, $\mathcal{\gamma}_{20}$, $\mathcal{\gamma}_{10}$) is the energy relaxation rate, and $\kappa_k$ is the single photon loss rate of the cavity $\mathrm{C}_k$. For simplicity, we set $\mathcal{\gamma}_{ff}=\Gamma_{d}$, $\mathcal{\gamma}_{4i}=\Gamma_{\gamma}/4,\mathcal{\gamma}_{3j}=\Gamma_{\gamma}/3$, $\mathcal{\gamma}_{20}=\Gamma_{\gamma},\mathcal{\gamma}_{10}=\Gamma_{\gamma}$, and $\kappa_k=\Gamma_{\kappa}$ in following numerical simulations.

\begin{figure}
  \centering
  \subfigure{\scalebox{0.6}{\includegraphics{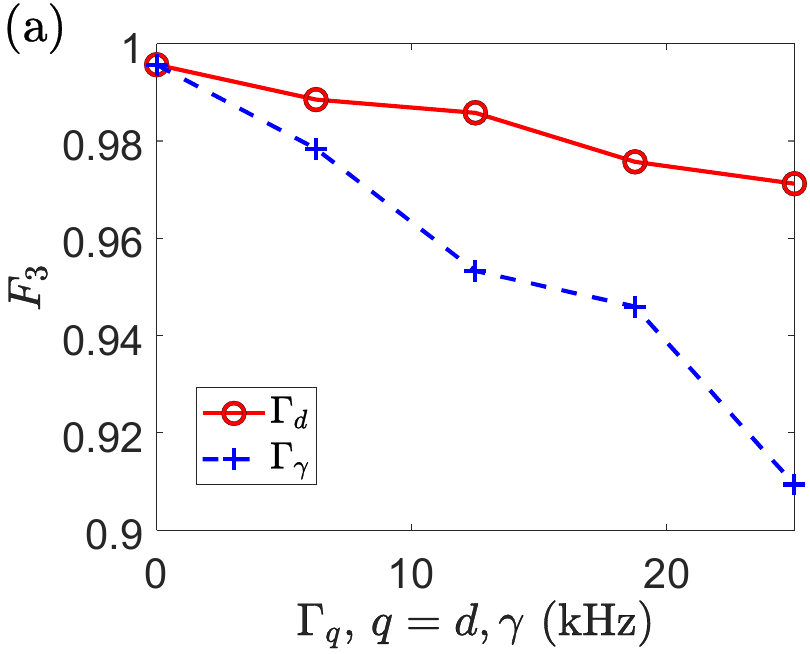}} \label{bite}}
  \subfigure{\scalebox{0.6}{\includegraphics{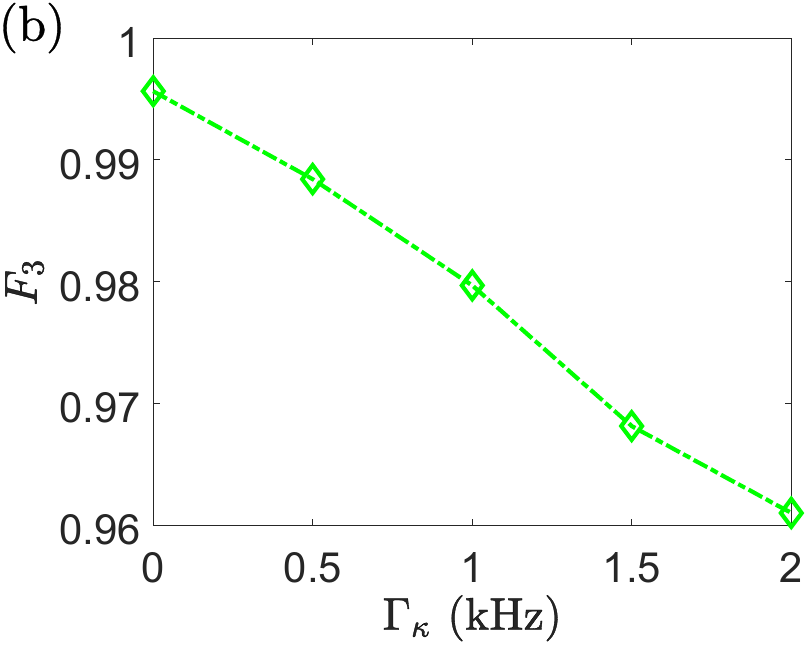}}\label{kappa} }
  \caption{(a) Fidelity $F_3$ versus the dephasing rate $\Gamma_{\Gamma}\in[0, 25]$ kHz or the qudit spontaneous emission rate $\Gamma_{\gamma}\in[0, 25]$ kHz, ($q=d,\gamma$). (b) Fidelity $F_3$ versus the single photon loss rate $\Gamma_{\kappa}\in[0, 2]$ kHz.}\label{dec}
\end{figure}

Based on the master equation in Eq.~(\ref{mas}), we examine the impact the dephasing, the energy relaxation, and the single photon loss.
In Fig.~\ref{dec}, the fidelity $F_3$ versus the dephasing rate $\Gamma_{d}$, the energy relaxation rate $\Gamma_{\gamma}$, and the single photon loss rate $\Gamma_{\kappa}$ of both cavities
are plotted. As shown in Fig.~\ref{bite}, the qudit energy relaxation has a greater impact compared to the qudit dephasing. The red-solid (blue-dashed) curve demonstrate that the fidelity $F\geq97\%$ ($F\geq91\%$) can be obtained with $\Gamma_{d}$ ($\Gamma_{\gamma}$) in the range $[0,25]$~kHz. 
In addition, Fig.~\ref{kappa} shows that the protocol can achieve the fidelity $F_3\geq 96\%$ for $\Gamma_{\kappa}\in[0,2]$ kHz. This suggests that the single photon loss is the main decoherence factor affecting the protocol, and cavities with higher quality factors are desired.

\textcolor{blue}{In superconducting system, the relaxation time $T_1$ and the dephasing time $T_2$ of flux qubits about 40--60~$\upmu$s (corresponding to the relaxation and dephasing rates about 16.67--25~kHz) have been previously reported \cite{Yan2016,5136262}. Here, it is worth noting that due to the large detuning condition ($\Delta_k\gg\lambda_k, k=1,2$), the qubit relaxation time and dephasing time are not significantly affected by the strong coupling.~For a detailed analysis, please refer to the appendix~\ref{appenC}.} Accordingly, the considered range $[0,25]$~kHz for the relaxation and dephasing rates in Fig.~\ref{bite} is reasonable and available for flux qubits. Moreover, the high quality factor $Q_c\sim3.5\times10^7$ to $Q_c\sim3\times10^9$ has been demonstrated for 3D microwave cavities \cite{science.aaf2941,PhysRevB.94.014506,PRXQuantum.4.030336}. Substituting the cavity frequencies $\omega_1=2\pi\times11.0346$ GHz and $\omega_2=2\pi\times4.0346$ GHz of the cavities $\mathrm{C}_1$ and $\mathrm{C}_2$, the ranges of single photon loss rate for the cavities $\mathrm{C}_1$ and $\mathrm{C}_2$ range in $[0.023,1.975]$~kHz and $[0.008,0.718]$~kHz, respectively. Therefore, the maximal single photon loss rate 2~kHz considered in Fig.~\ref{kappa} is reasonable. According to the numerical results in this section, the protocol can still generate NOON states with relatively high fidelity in the presence of decoherence with current technology in superconducting systems.

\textcolor{blue}{In above Sec.~\ref{Ab} to Sec.~\ref{Ac}, we have evaluated the impact of various disturbing factors on the protocol in isolation. However, in reality, these factors often work together. Therefore, building upon current experimental conditions and some theoretical works~\cite{Yan2016,5136262,science.aaf2941,PhysRevA.107.013702,PhysRevA.109.042615,Zhao2021,PhysRevA.76.042319,PhysRevA.105.062436,PhysRevB.94.014506,PRXQuantum.4.030336}, we now assess the performance of the protocol under all disturbing factors. Table \ref{tab2} presents the fidelity $F_3$ of the target NOON states under all disturbing factors. From Table \ref{tab2}, we can know that the protocol maintains acceptable fidelity ($>90\%$) under the current experimental conditions~\cite{Yan2016,5136262,science.aaf2941,PhysRevA.76.042319,PhysRevA.105.062436,PhysRevB.94.014506,PRXQuantum.4.030336}. This shows that the protocol has good experimental feasibility.}

\section{CONCLUSION}\label{V}

In this manuscript, we propose a protocol for generating the NOON states in a superconducting system consisting of two microwave cavities coupled to a five-level qudit. By appropriately engineering the couplings between the qudit and the cavities, as well as applying tailored driving fields to the qudit, the NOON states can be prepared through a three-step operation. The qudit is first excited from its lowest energy level to a superposition state of two higher levels. Then, utilizing effective single-photon drives that are conditional on the state of the qudit, the cavities selectively evolve into coherent states with an average photon number $N$. Finally, frequency-matched pulses are applied to drive the qudit back to its lowest level, disentangling it from the cavities, leaving the two cavities in the target NOON states with photon number $N$. For the first and third steps, the control fields are designed using reverse engineering and optimal control techniques, which significantly enhance the robustness against potential control errors.

Through numerical simulations, we investigate the effects of the systematic errors for the control field amplitudes and the deviation of the coupling strengths, the crosstalk between both cavities, and decoherence on the performance of the protocol. The results demonstrate that the protocol can achieve acceptable fidelity for the NOON states even in the presence of these disturbances. Here, the NOON states is generated in three steps by adjusting only external classical fields without requiring complex nonlinear interactions. Meanwhile, the qudit-cavity coupling strengths and the qudit level spacings is fixed, which is friendly to experimental realization. Therefore, the protocol may also be applicable for generating the NOON states in a wide range of physical systems.

\begin{acknowledgments}
Y.-H. K. was supported by the National Key Research and Development Program of China under Grant No. 2024YFA1408900, National Natural Science Foundation of China under Grant No. U21A20436. Y.-H.C. was supported by the National Natural Science Foundation of China under Grant No. 12304390 and 12574386, the Fujian 100 Talents Program, and the Fujian Minjiang Scholar Program. Y. Xia was supported by the National Natural Science Foundation of China under Grant No. 62471143, the Key Program of National Natural Science Foundation of Fujian Province under Grant No. 2024J02008, and the project from Fuzhou University under Grant No. JG2020001-2. Z.-C.~S. was supported by the National Natural Science Foundation of China under Grant No.~62571129 and the Natural Science Foundation of Fujian Province under Grant No.~2025J01456.
\end{acknowledgments}

\begin{appendix}

\begin{figure*}
  \centering
  \includegraphics[width=\linewidth]{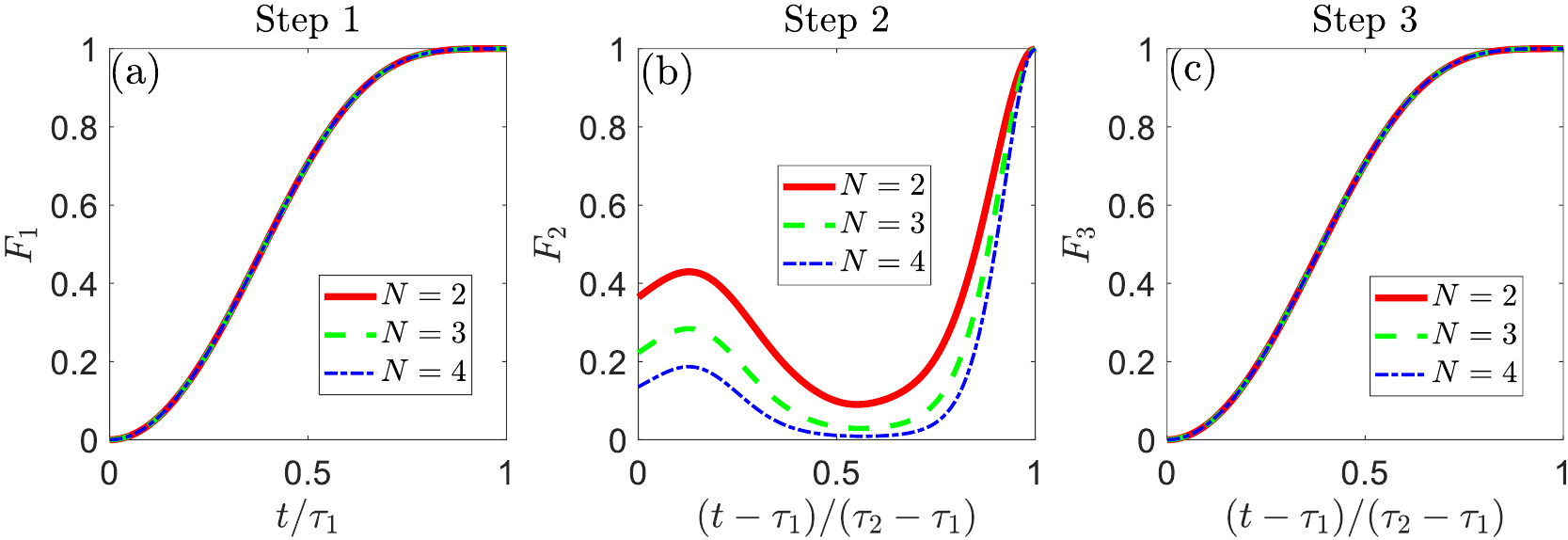}\caption{\textcolor{blue}{Fidelity $F_p$ of the target state in Step $p$ ($p=1,2,3$) versus $t$ for the photon number $N=2,3,4$, where the system evolution is governed by the effective Hamiltonian $H_e$.}}\label{effstep}
\end{figure*}

\begin{figure}
  \centering
  \includegraphics[width=8cm]{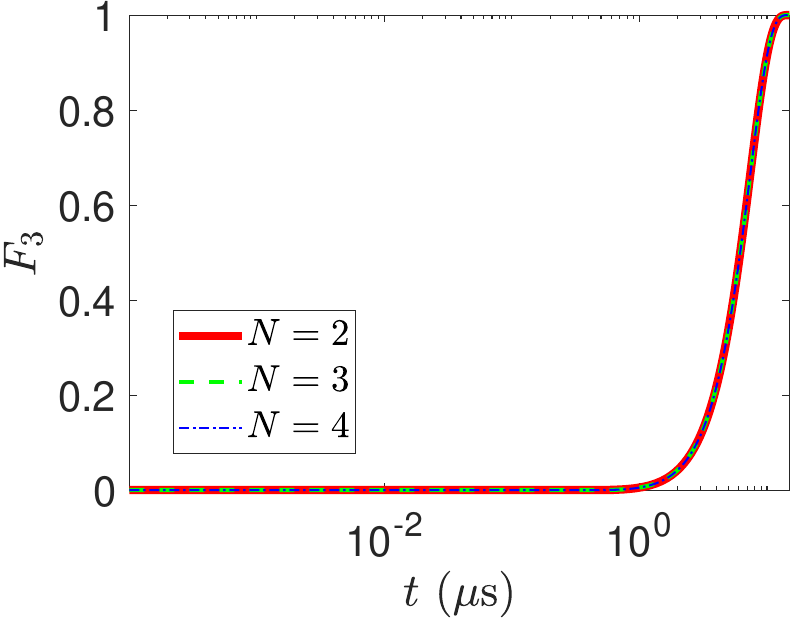}\caption{\textcolor{blue}{Fidelity $F_3$ versus $t$ for the photon number $N=2,3,4$, where the system evolution governed by the effective Hamiltonian $H_e$.}}\label{eff234}
\end{figure}

\section{Design of the optimized pulse}\label{appenA}
To begin with the invariant-based reverse engineering, we first rewrite the Hamiltonian $H_1^{\mathrm{eff}}(t)$ in Eq.~(\ref{e3}) as
\begin{eqnarray}\label{e5}
H_1^{\mathrm{eff}}(t)&=&\mathrm{Re}[\Omega_{s_1}(t)]\sigma_x+\mathrm{Im}[\Omega_{s_1}(t)]\sigma_y+0\times\sigma_z,\
\end{eqnarray}
where the Pauli operators are defined as
\begin{eqnarray}
\sigma_x&=&\vert\xi_1\rangle\langle\xi_2\vert+\vert\xi_2\rangle\langle\xi_1\vert,\cr\cr
\sigma_y&=&-i\vert\xi_1\rangle\langle\xi_2\vert+i\vert\xi_2\rangle\langle\xi_1\vert,\cr\cr
\sigma_z&=&\vert\xi_1\rangle\langle\xi_1\vert-\vert\xi_2\rangle\langle\xi_2\vert,
\end{eqnarray}
with $\vert\xi_1\rangle=|0,0,0\rangle$ and $\vert\xi_2\rangle=|\Phi_+\rangle$.
Here, the Pauli operators satisfy the commutation relations
\begin{eqnarray}
[\sigma_x,\sigma_y]=2i\sigma_z,\ [\sigma_y,\sigma_z]=2i\sigma_x,\
[\sigma_z,\sigma_x]=2i\sigma_y.
\end{eqnarray}

To study the evolution of the two-level system by invariant-based reverse engineering \cite{CXPRA83_2011,CXPRA86_2012,PhysRevA.109.062610,PhysRevResearch.4.013233}, we need find a dynamical invariant $I(t)$ fulfilling \cite{LewisJMP10_1969}
\begin{eqnarray}\label{e8}
i\frac{\partial}{\partial t}I(t)-[H_1^{\mathrm{eff}}(t), I(t)]=0.
\end{eqnarray}
Since the Hamiltonian in Eq.~(\ref{e5}) has a SU(2) dynamical structure, a dynamical invariant is found as \cite{TorronteguiPRA89_2014}
\begin{eqnarray}
I(t)&=&\sin\theta\sin\beta~\sigma_x+\sin\theta\cos\beta~\sigma_y+\cos\theta~\sigma_z,
\end{eqnarray}
where $\theta$ and $\beta$ are two time-dependent parameters.

By reversely solving Eq.~(\ref{e8}), the expressions of $\mathrm{Re}[\Omega_{s_1}(t)]$ and $\mathrm{Im}[\Omega_{s_1}(t)]$ are given by
\begin{eqnarray}\label{e10}
\mathrm{Re}[\Omega_{s_1}(t)]&=&\left(\dot{\beta}\tan\theta\sin\beta-\dot{\theta}\cos\beta\right)/2,\cr\cr
\mathrm{Im}[\Omega_{s_1}(t)]&=&\left(\dot{\beta}\tan\theta\cos\beta+\dot{\theta}\sin\beta\right)/2.
\end{eqnarray}
The eigenvectors of the dynamical invariant $I(t)$ are
\begin{eqnarray}
\vert\zeta_+(t)\rangle&=&e^{i\beta/2}\cos\frac{\theta}{2}\vert\xi_2\rangle+ie^{-i\beta/2}\sin\frac{\theta}{2}\vert\xi_1\rangle, \cr\cr
\vert\zeta_-(t)\rangle&=&ie^{i\beta/2}\sin\frac{\theta}{2}\vert\xi_2\rangle+e^{-i\beta/2}\cos\frac{\theta}{2}\vert\xi_1\rangle,
\end{eqnarray}
with the eigenvalues $1$ and $-1$, respectively. Using the eigenvectors $|\zeta_\pm(t)\rangle$, two orthogonal solutions $|\psi_\pm(t)\rangle$ of the Schr\"{o}dinger equation $i|\dot{\psi}(t)\rangle=H_1^{\mathrm{eff}}(t)|\psi(t)\rangle$ are derived as
\begin{eqnarray}
|\psi_\pm(t)\rangle&=&e^{i\mu_\pm(t)}\vert\zeta_\pm(t)\rangle,
\end{eqnarray}
where $\mu_{\pm}(t)$ are the Lewis-Riesenfeld (LR) phases given by \cite{LewisJMP10_1969}
\begin{eqnarray}
\mu_{\pm}(t)&=&\langle\zeta_{\pm}(t)|[i\frac{\partial}{\partial t}-H_1^{\mathrm{eff}}(t)]|\zeta_{\pm}(t)\rangle \cr\cr
&=&\pm\int_0^t\frac{\dot{\beta}(t')}{2\cos[\theta(t')]}dt'.
\end{eqnarray}
Here, we select the evolution paths $\vert\psi_+(t)\rangle$ to realize the evolution $|\psi_+(0)\rangle=|\xi_1\rangle\rightarrow|\psi_+(T)\rangle=e^{i\chi(T)/2}|\xi_2\rangle$. Then, the corresponding boundary conditions can be specified as
\begin{eqnarray}\label{e14}
\theta(0)=\beta(0)=\beta(\tau_1)=0,\ \theta(\tau_1)=\pi.
\end{eqnarray}

So far, we have obtained the expressions of the classical fields $\Omega_{s_1}(t)$ by using reverse engineering. Next, to enhance the robustness against the systematic errors in the control function $\Omega_{s_1}(t)$, we need to further design the control parameters $\theta$ and $\beta$ by nullifying systematic-error sensitivity \cite{Ruschhaupt_2012}. Considering the systematic errors of the control function $\Omega_{s_1}(t)$ with the systematic error rate $\delta$, i.e., $\Omega_{s_1}(t)\rightarrow(1+\delta)\Omega_{s_1}(t)$, the effective Hamiltonian in Eq.~(\ref{e5}) becomes $H_s(t)=(1+\delta)H_1^{\mathrm{eff}}(t)$.
By using the time-dependent perturbation theory, one gets \cite{Ruschhaupt_2012}
\begin{eqnarray}
\vert\psi_{\delta}(T)\rangle&=&\vert\psi(T)\rangle-i\delta\int_0^{\tau_1}dtU_0(\tau_1,t)H_1^{\mathrm{eff}}(t)\vert\psi(t)\rangle\cr&+&O(\delta^2),
\end{eqnarray}
where $U_0(\tau_1,t)=\sum_{l=\pm}|\psi_l(\tau_1)\rangle\langle\psi_l(t)|$ is the unperturbed time evolution operator for the time interval $[t,\tau_1]$, $O(\delta^2)$ are the terms with orders equal to or greater than $\delta^2$, and $\ket{\psi_{\delta}(t)}$ ($\ket{\psi(t)}$) is the state of the system with (without) the influence of the systematic errors.

The fidelity of the evolution $|\xi_1\rangle\rightarrow|\xi_2\rangle$ along the path $|\psi_+(t)\rangle$ can be approximately calculated as
\begin{eqnarray}
F&\approx&1-\delta^2\left|\int_0^{\tau_1}\langle\psi_-(t)\vert H_1^{\mathrm{eff}}(t)\vert\psi_+(t)\rangle dt\right|^2 \cr\cr
 &=&1-\delta^2\left|\int_0^{\tau_1}e^{i\chi(t)}\dot{\theta}\sin^2\theta dt\right|^2,
\end{eqnarray}
with $\chi(t)=2\mu_+(t)$.
According to Refs.~\cite{Ruschhaupt_2012,PhysRevA.97.062317}, the systematic-error sensitivity can be calculated by
\begin{eqnarray}
Q=-\frac{1}{2}\frac{\partial^2F}{\partial\delta^2}\mid_{\delta=0}
=\left|\int_0^{\tau_1}e^{i\chi(t)}\dot{\theta}\sin^2\theta dt\right|^2.
\end{eqnarray}

To nullify the systematic-error sensitivity, we assume $\chi(t)\!=\!A(2\theta-\sin2\theta)$ with a time-independent coefficient $A$ \cite{PhysRevLett.111.050404}. Then, the systematic-error sensitivity is derived as
\begin{eqnarray}
Q=\sin^2(A\pi)/A^2,
\end{eqnarray}
which implies that $Q=0$ can be obtained for $A=\pm1,\pm2,\cdots$. Here, $A=1$ is chosen for simplicity. Using the relation $\chi(t)=2\mu_+(t)$, one derives
\begin{eqnarray}
\dot{\chi}(t)&=&2\dot{\theta}(1-\cos2\theta)=\dot{\beta}/\cos\theta,\cr\cr
\Rightarrow\dot{\beta}(t)&=&2\dot{\theta}\cos\theta(1-\cos2\theta).
\end{eqnarray}
According to the boundary conditions in Eq.~(\ref{e14}), the expression of $\beta$ is given by
\begin{eqnarray}\label{e18}
\beta(t)&=&\frac{4}{3}\sin^3\theta.
\end{eqnarray}
Substituting $\beta(t)$ in Eq.~(\ref{e18}) into Eq.~(\ref{e10}), the specific expressions for the control fields can be obtained as
\begin{eqnarray}
\mathrm{Re}[\Omega_{s_1}(t)]&=&\frac{\dot{\theta}}{2}\left(4\sin\beta\sin^3\theta-\cos\beta\right),\cr\cr
\mathrm{Im}[\Omega_{s_1}(t)]&=&\frac{\dot{\theta}}{2}\left(4\cos\beta\sin^3\theta+\sin\beta\right).
\end{eqnarray}
In addition, based on the boundary conditions in Eq.~(\ref{e14}), the parameter $\theta(t)$ can be selected as
\begin{eqnarray}\label{27}
\theta(t)&=&\pi\sin^2(\pi t/2\tau_1).
\end{eqnarray}
At this point, the expression of $\Omega_{s_1}(t)$ is obtained.

\section{Validation of the effective process}\label{appenB}

In this section, the system evolution under the control of an effective Hamiltonian $H_e$ is investigated, where $H_e=H_1^{\mathrm{eff}}, H_2^{\mathrm{eff}}, \mathcal{H}_{3}^{\mathrm{eff}}$ for Step 1,2,3, respectively. In this case, the system is governed by the equation
 \begin{eqnarray}\label{A32}
\dot{\rho}(t)&=&-i[H_t,\rho(t)].
\end{eqnarray}
For the different photon number $N=2,3,4$, we examine the fidelity $F_p$ ($p=1,2,3$) of the target state versus $t$ at each step, as well as the fidelity $F_3$ of the desired NOON states versus $t$ throughout the entire evolution process, as illustrated in Figs. \ref{effstep} and \ref{eff234}. Obviously, for the different photon number $N=2,3,4$, the fidelity $F_p$ reaches exactly 1 after each step and the fidelity $F_3$ of the desired NOON states reaches 1 after three steps. It is clear that the NOON states can be prepared precisely under the control of the effective Hamiltonian $H_e$. This confirms the validity of the effective Hamiltonian $H_e$. However, since the reduction of the original Hamiltonian $H$ to the effective Hamiltonian $H_e$ is a second-order approximation process, some higher-order terms and non-dominant terms are neglected in this process. Therefore, the fidelity $F_3$ of the desired NOON states cannot be exactly reached to 1 under the control of the original Hamiltonian $H$.


\section{The influence of strong coupling on qubit relaxation time and dephasing time}\label{appenC}
\textcolor{blue}{When the coupling strength reaches hundreds of megahertz, the longitudinal relaxation time $T_1$ should be calculated by considering both direct qubit relaxation and cavity-mediated relaxation.~Similarly, for the calculation of the transverse relaxation time $T_2$, three components must be taken into account: dephasing induced by longitudinal relaxation, intrinsic dephasing of the qubit itself, and phase noise induced by strong coupling. Based on the above conclusions, the expressions for $T_1$ and $T_2$ are revised as follows \cite{PhysRevB.94.014506,RevModPhys.93.025005}: (to distinguish them from $T_1$ and $T_2$ in the main text, we use $T_1'$ and $T_2'$ to represent them)
\begin{align}\label{Aa1}
T_1'&=1/\Gamma_1',\ \ \ \ \Gamma_1'\approx\left(\frac{\lambda_1^2}{\Delta_1^2}+\frac{\lambda_2^2}{\Delta_2^2}\right)\Gamma_{\kappa}+\Gamma_{\gamma},\cr
T_2'&=1/(\frac{1}{2T_1'}+\Gamma_{\phi}),\ \ \ \Gamma_{\phi}\approx\left(\frac{\lambda_1^2}{\Delta_1^2}+\frac{\lambda_2^2}{\Delta_2^2}\right)\kappa_{\phi}+\Gamma_{d},\ \ \
\end{align}
where $\Gamma_k$, $\Gamma_{\gamma}$, and $\Gamma_d$ are the single photon loss rate of the cavity, the energy relaxation rate of the qubit, and the dephasing rate of the qubit in main text of the manuscript, respectively. It is noted that $\kappa_{\phi}$ the intrinsic dephasing rate of the cavity. For a cavity with high quality factor ($Q_c\sim7\times10^7$), the interference factor is usually negligible ($\kappa_{\phi}/2\pi\leq40$ Hz) \cite{PhysRevB.94.014506}.
In Eq. (\ref{Aa1}), the loss terms $\left(\frac{\lambda_1^2}{\Delta_1^2}+\frac{\lambda_2^2}{\Delta_2^2}\right)\Gamma_{\kappa}$ and $\left(\frac{\lambda_1^2}{\Delta_1^2}+\frac{\lambda_2^2}{\Delta_2^2}\right)\kappa_{\phi}$ here arise from the coupling between the cavity and qubit. When the coupling strength is large enough, these terms are not negligible. However, since the protocol operates under large detuning condition (where $\Delta_k\gg\lambda_k, k=1,2$), the loss caused by strong couplings can be neglected despite the coupling strength reaching hundreds of megahertz. Consequently, $T_1'\approx T_1$ and $T_2'\approx T_2$.}

\end{appendix}

\bibliographystyle{apsrev4-1}
\bibliography{reference}

\end{document}